\input epsf
\def\beq{\begin{equation}}
\def\eeq{\end{equation}}
\def\bey{\begin{eqnarray}}
\def\eey{\end{eqnarray}}
\def\kms{\mbox{\rm \,km\,s}^{-1}}
\def\muas{\mu{\rm as}}
\def\muasyr{\mu{\rm as\, yr}^{-1}}
\documentstyle[epsfig,twoside,rotating,referee]{aa}
\begin{document}
\title
{Evolution of the Galactic Potential and Halo Streamers
with Future Astrometric Satellites}
\author
{HongSheng Zhao\inst{1},
Kathryn V. Johnston\inst{2}, 
Lars Hernquist\inst{3},
David N. Spergel\inst{4}}
\institute{
Sterrewacht Leiden, Niels Bohrweg 2, 2333 CA,
Leiden, The Netherlands
\and
Institute for Advanced Study, Princeton, NJ 08450
\and
Board of Studies in Astronomy and Astrophysics,
University of California, Santa Cruz, CA 95064
\and
Princeton University Observatory, Princeton University,
Princeton, NJ 08540
}

\offprints{Zhao: hsz@strw.leidenuniv.nl}
\date{Received, accepted }
\thesaurus{}

\maketitle

\markboth{Zhao et al. : Evolution of the Galactic Potential and Halo Streamers}
{Zhao et al. : Evolution of the Galactic Potential and Halo Streamers}



\begin{abstract}
ESA's Global Astrometric Interferometer for Astrophysics (GAIA) holds
the promise of mapping out the detailed phase space structure of the
Galactic halo by providing unprecedented annual proper motion and
parallax of $1-10\muas$ astrometric accuracy (Gilmore et al. 1998).
Unlike NASA's Space Interferometry Mission (SIM), which will achieve
similar accuracies but is a pointed instrument, GAIA will be able to
construct a global catalogue of the halo.  Here we study proper
motions of giant branch stars in a tidal debris torn from a small
satellite (a $10^{5-7}L_\odot$ Galactic dwarf galaxy or globular
cluster) in the halo.  We follow the evolution of a cold (velocity
dispersion of 10 km/s) stream on a nearby (between 8-50 kpc) polar
orbit in a variety of histories of the potential of the Galaxy, and
observe the bright ($V<18$mag) members of the debris tail with GAIA
accuracy.  We simulate effects due to the growing or flipping of the
Galactic disk over the past 4 Gyrs or the perturbation from a massive
accreted lump such as the progenitor of the Magellanic Clouds.
Our simulations suggest that
the results of Johnston, Zhao, Spergel \& Hernquist (1999) and Helmi, Zhao
\& de Zeeuw (1999) for static Galactic potentials are likely to be
largely generalizable to realistic time-dependent potentials:
a tidal debris remains cold in spite of evolution
and non-axial symmetry of the potential.
GAIA proper motion measurements of debris stars 
might be used to probe both Galactic
structure and Galactic history.  We also 
study several other factors influencing our ability to identify streams,
including accuracy of radial velocity
and parallax data from GAIA, and contamination from random field stars.  
We conclude that nearby, cold streams could be
detected with GAIA if these cousins of the Sagittarius stream exist.
\end{abstract}

\section{Introduction}

Tidal streams in the Galactic halo are a natural prediction of
hierarchical galaxy formation, where the Galaxy builds up its mass by
accreting smaller infalling galaxies.  They are often traced by
luminous horizontal and giant branch (HB and GB) stars outside the
tidal radius the satellite, by which we mean either a dwarf galaxy or
a globular cluster in the Galactic potential.  These extra-tidal stars
have been seen for the Sagittarius dwarf galaxy (Ibata, Gilmore, \&
Irwin 1994) and for globular clusters (cf. Grillmair et al. 1998, Irwin
\& Hatzidimitriou 1995) as a result of tidal stripping, shocking or
evaporation.  That extra-tidal material (stars or gas clouds) traces
the orbit of the satellite or globular cluster has long been known to
be a powerful probe of the potential of the Galaxy in the halo.  This
technique has been exploited extensively particularly in the case of
the Magellanic Clouds and Magellanic Stream (Murai \& Fujimoto 1980,
Putman et al. 1999) and the Sagittarius dwarf galaxy (Ibata, Gilmore
\& Irwin 1995, Ibata, Wyse, Gilmore \& Suntzeff 1997, Zhao 1998 and
references therein).

Helmi, Zhao \& de Zeeuw show that streams can be
identified by as peaks in the distribution in the angular momentum
space, measurable with GAIA.  Once identified, we can fit a stream
with an orbit or more accurately a simulated stream in a given
potential.  Johnston, Zhao, Spergel \& Hernquist (1999) show that a
few percent precision in the rotation curve, flattening and
triaxiality of the halo is reachable by mapping out the proper motions
(with SIM accuracy) and radial velocities along a tidal stream $\> 20$
kpc from the Sun.  In particular, they show that the fairly large error
in distance measurements to outer halo stars presents no serious
problem since one can predict distances theoretically using the known
narrow distribution of the angular momentum or energy along the tails
associated with a particular Galactic satellite.  We expect these
results should largely hold for streams detectable by GAIA.  These
numerical simulations are very encouraging since they show that it is
plausible to a learn great deal about the Galactic potential with even
a small sample of stream stars from GAIA.  Some unaddressed issues
include whether stream members will still be identifiable in angular
momentum in potentials without axial symmetry, and the robustness of
both methods if the Galactic potential evolves in time.

Here we illustrate the effects of including a realistic evolution
history of the Galaxy's potential.  The simulations are observed with
GAIA accuracy.  We discuss whether bright stars in a stream might
still be identified using the 6D information from GAIA.

\section{Streams in realistic time-varying Galactic potentials}

\subsection{Evolution of the Galactic potential}
To simulate the effect of the evolution and flattening of the
potential on a stream, we will consider a satellite orbiting in the following 
simple, flattened, singular isothermal potential $\Phi(r,\theta,t)$
\beq
\Phi_G(r,\theta,t) = V_0^2 \left[A_s \log r + 
{\epsilon \over 2} \cos 2\theta \right],
\eeq
where 
\beq
A_s(t)=1-\epsilon_0+\epsilon(t),
~~\epsilon(t)=\epsilon_0 \cos{2\pi t \over T_G}.
\eeq
This model simulates the effect that the Galaxy becomes
more massive and flattened in potential as it grows a disc.
It is time-dependent but maintains a rigorously flat rotation curve
at all radii, where $(r,\theta)$ are the spherical coordinates
describing the radius and the angle from the North Galactic Pole, $t$
is defined such that $t=0$ would be the present epoch.
The time-evolution is such that 
the Galactic potential grows from prolate at time $t=-T_G/2$
to spherical at time $t=-T_G/4$, and then to oblate at $t=0$;
a more general prescription of the temporal variation 
might include a full set of Fourier terms.  

We adopt parameters 
\beq
V_0=200\kms, ~~~ \epsilon_0=0.1, 
\eeq
for the present-day circular velocity, and the flattening
of the equal potential contour of the Galaxy respectively.  
A small $\epsilon_0$ guarantees
a positive-definate volume density of the model everywhere at all time.
We set $T_G/4=4$Gyr, 
a reasonable time scale for the growth of the Galactic disc.

We have also considered Galactic potentials with a flipping disc and
with a massive perturber (Zhao et al. 1999).  The results are
qualitatively the same.  In the following we will illustrate our
points with the potential with a growing disc $\Phi_G$.

\subsection{Evolution of the satellite}

Following Helmi \& White (1999), we assume that the particles in the
disrupted satellite are initially distributed with an isotropic
Gaussian density and velocity profile with dispersions $0.4$ kpc and
$4\kms$ respectively.  These particles are released instantaneously
at the pericenter $8$ kpc from the center $4$ Gyrs ago.  These
parameters might be most relevant for satellites such as the
progenitor of the Sagittarius stream.  We simulate observations of
mock data of 100 bright Horizontal Branch stars convolved with
GAIA accuracy.

Johnston (1998) showed that a satellite on rosette orbits in the outer
halo $\ge 20$kpc leaves behind a nice thin spaghetti-like tail.  Helmi
\& White (1999) put their satellites on plunging orbits which come
within the solar circle, and found that they become colder in velocity
space, but very mixed in the coordinate space after evolving for a
Hubble time.  While the cold, linear structures seen in Johnston's
simulations are ideal for studying the potential, these stars at large
distance are likely too faint for GAIA except for those high up on the
tip of the giant branch.

Our choice of the pericenter of the satellite is somewhere in between
previous authors.  We concentrate on satellites which fall in and are
disrupted recently (about 4 Gyrs ago, well after the violent
relaxation phase) and maintain a cold spaghetti-like structure.  We
put the satellites on relatively tight orbits but which lie outside
the solar circle (pericenter of about 8 kpc and apocenter of about 40
kpc) such that the bright member stars in the stream are still within
the reach of detectability of GAIA.  Such streams typically go around
the Galaxy less than 5 times since disruption, and are typically far
from fully phase-mixed.

Fig.1 shows the orbit and morphology of the simulated stream in the
potential described in \S 2.1.  
The orbit of the disrupted satellite is chosen so that the
released stream stays in the polar $xz$ plane, which passes through
the location of the Sun and the Galactic center; the $xyz$ coordinate
is defined such that the Sun is at $x=-8$ kpc and $y=z=0$.

Fig. 2 shows the simulated streams in energy and angular momentum
space.  By and large the energy $E$ of
particles across each stream is spread out only in a narrow range at
each epoch in the three models; the same holds but to a lesser extent
for the angular momentum vector ${\bf J}$.  This implies that stars in
the stream are largely coeval even in the presence of realistic,
moderate evolution of the Galactic potential.  
The energy and angular
momentum is also modulated with particle position in a sinusoidal way
across the stream, an effect which in principle can be used to infer
the evolution rate of the Galactic potential and the flattening of the
potential.  

\subsection{Some analytical arguments}

To understand the above results analytically,
let's follow the energy evolution of two particles (1 and 2)
in an extremely simplified evolution history of the Galactic potential
\beq
\Phi(r,\theta,t) = V_0^2 \log r - g|z| {t+t_G \over t_G},
\eeq
where the potential varies linearly on a time scale $t_G$
from spherical at time $t=-t_G$ to a present-day $t=0$ 
flattened potential with a uniform razor thin disk in the $z=0$ plane;
$g$ is the surface gravity of the disk.
Assume the two particles are released from the satellite with a slight
initial energy difference $\Delta_i$, which causes them to drift apart 
in orbital phase.  
Given enough time the initial phase-bunching of the stream particles should 
all be dispersed away as a stream can develop tails that wrap around the sky.
More generally, we let
$t_{\rm ph}(\Delta_i)$ be a timescale for the two particles
to drift sufficiently apart and become out of phase with each other.
The two particles' energy difference $(E_2-E_1)$ 
at the present epoch $t=0$ is then readily computed from
\beq
(E_2-E_1)=
\Delta_i+\int_{-t_G}^0 \!\! dt {\partial \over \partial t} (\Phi_2-\Phi_1)
=\Delta_i+\left({t_{\rm ph} \over t_G}\right) \xi g z_{\rm max},
\eeq
where $z_{\rm max}$ is a typical scale height for the orbit, 
and the dimensionless factor
\beq
\xi \equiv 
{1 \over t_{\rm ph}} \int_{-t_G}^0\!\! dt \left( \eta_1-\eta_2 \right),
~~~0 \le \eta \equiv {|z|\over z_{\rm max}} \le 1.
\eeq

Interestingly the factor $\xi$ is of order unity if not smaller.  This
is because $\eta_1-\eta_2$ should take any value between -1 and 1 with
equal chance once the two particles are completely out of phase.
Hence, we are integrating over an oscillatory function of $t$ in the
range $-t_G + t_{\rm ph} \ge t \ge 0$.  This limits $|\xi|$ at the
upper end.

The above equations suggest that:
(i) The spread of energy among particles
is proportional to the change of the potential, here $g$ of the disk:
particles stay on the same orbit for a static potential.
(ii) If the change of the potential is abrupt, e.g., like a step-function,
then the energy spread $(E_2-E_1) \sim \Delta_i + g \left[|z_2|-|z_1|\right]$ 
is roughly proportional to the separation of the particles at the time of
change.  Adjacent particles with $(z_1-z_2) \ll z_{\rm max}$
should also be adjacent in the energy space.
This kind of sudden change of potential 
might be relevant for galaxy formation models with minor infalls
on the orbital time scale. 
(iii) If the potential grows slowly, 
e.g., adiabaticly with $t_G \rightarrow \infty$, 
then we expect only a very small spread in the energy space, no greater than 
$|\Delta_i|+\left({t_{\rm ph} \over t_G}\right) \left(g z_{\rm max}\right)$.
This is consistent with the argument that
energy differences between initially similar orbits remain
small because of adiabatic invariance of the actions of orbits
(as explicitly shown in the spherical case
by Lynden-Bell \& Lynden-Bell 1995).

\subsection{Accuracy of parallax and radial velocity from GAIA}
Fig. 3 shows the stream in various slices of the 6D phase space
observable by GAIA.
One of the challenges of using streams to constrain the potential is
the measurement errors of proper motion ($\mu$ in $\muasyr$),
parallax ($\pi$ in $\muas$), and heliocentric radial velocity ($V_h$
in $\kms$), particularly of the latter two.
For HB stars, the errors are functions of parallax.  
We find that this simple formula
\beq\label{error}
{\rm Err}[\pi]=1.6{\rm Err}[\mu]={\rm Err}[V_h]=5+50 (50/\pi)^{1.5},
\eeq
works well in approximating 
GAIA specifications (Lindegren 1998, private communications) 
on errors on parallax, proper motion and radial velocity.

Proper motions will be precisely measured everywhere in the Galaxy.  A
remarkable feature of GAIA is that it can resolve the internal proper
motion dispersion of a satellite.  Even a small satellite (e.g. a
globular cluster) has a dispersion $\sigma \ge 4 \kms$.  This means
that the dispersion in the proper motion is larger than the resolution
limit of GAIA for HB stars anywhere within $20$kpc ($\pi \ge 50\muas$,
${\rm Err}[\mu] \le 40\mu{\rm as/yr}=4\kms/20{\rm kpc}$,
cf. ~\ref{error}).  In comparison, radial velocities remain accurate
at $20\kms$ level, and astrometric parallaxes remain superior than
photometric parallaxes only within $10$ kpc ($\pi \ge 100\muas$, $V
\le 16$mag.)  because of rapid growth of error bars with magnitude of
a star.

Fortunately, a ``theoretical'' parallax and heliocentric velocity
can be predicted to a good accuracy 
from the property that the angular momentum and 
energy are roughly constant, i.e.,
\beq
{\bf J} = {\bf r \times V} \sim {\rm constant},
\eeq
\beq
E \sim (200\kms)^2 \log r + {1 \over 2} {\bf V}^2 \sim {\rm constant}.
\eeq
Here we pretend that the Galactic potential is the simplest
spherical, static and isothermal potential, and feed in the accurately
measured proper motions.  Surprisingly, these
very rough approximations yield fairly accurate parallaxes
($\sim 10\%$) and heliocentric velocities ($\sim 30\kms$) as shown
by the narrow bands in Fig. 3; predictions tend to be poorer
for particles in the center and anti-center directions, where the
angular momentum ${\bf J}$ becomes insensitive to the heliocentric velocity.
But overall it appears promising to apply this method to
predict velocities and parallaxes for fainter stars,
where the predictions are 
comparable to or better than directly observable by GAIA.
The accuracy of these predictions is verifiable with direct
observations brighter members of a stream.

In essence our method is a variation of the classical method of
obtaining ``secular parallaxes''.  A polar stream with zero net
azimuthal angular momentum, $J_z \sim 0$, makes the simplest example.
Any proper motion in the longitude direction $\mu_l$ is merely due to
solar reflex motion, so parallaxes can be recovered to about $10\muas$
accuracy!) from the linear regression $\pi \sim |\mu_l|/40$ shown in
the top right panel of Fig. 3.

\subsection{Cool streams}

Fig. 3 and 4 show a somewhat surprising result: streams stay
identifiable in a variety of realistic time-dependent potentials.
Fig. 3. shows a slice of a stream in the proper motion vs. proper
motion diagram as well as the position-proper motion diagram.  A
stream shows up as a narrow feature, clearly traceable in the
position-proper motion diagram after convolving with GAIA errors.  The
narrowness of the distribution in the proper motion space argues that
stars in a stream are distinct from random field stars.

We have also run simulations with various parameters for the Galactic
potential and orbit and initial size of the satellite.  Fig. 4 shows a
few streams in the position vs. proper motion diagram.  While
extensive numerical investigations are clearly required to know
whether these ``cool'' linear features can be used to decipher the
exact evolution history, evolution itself clearly does not preclude
the identification of streams.  We caution that the structure of a
stream can become very noisy for highly eccentric orbits with a
pericenter smaller than 8 kpc and/or for potentials where the temporal
fluctuation of the rotation curve is greater than 10\%.  These noisy
structures as a result of strong evolution can be challenging to
detect.

We conclude that tidal streams are excellent tracers of the Galactic
potential as long as a stream maintains a cool, spaghetti-like
structure, in particular, the results of Johnston et al. (1999) and
Helmi et al. (1999) for static Galactic potentials are largely
generalizable to realisticly evolving potentials.  However,
perhaps the most exciting implication of these preliminary results is
that by mapping the proper motions along the debris with GAIA we could
eventually set limits on the rate of evolution of the Galactic
potential, and distinguish among scenarios of Galaxy formation.

\section{Discussions of Strategies}

\subsection{Why targeting streams?}

Which is the better tracer of the Galactic potential, stars in a cold
stream or random stars in the field?  Classical approaches use field
stars or random globular clusters or satellites as tracers of the
Galactic potential.  Assuming that they are random samples of the
distribution function (DF) of the halo, one uses, e.g., the Jeans
equation, to obtain the potential.  One need a large number of stars
to beat down statistical fluctuations, typically a few hundred stars
at each radius for ten different radii.  The problem is also often
under-constrained because of the large number of degrees of freedom in
choosing the 6-dimensional DF.  Another complication is that one
generally cannot make the assumption that the halo field stars are in
steady state as an ensemble because it typically takes much longer
than a Hubble time to phase-mix completely at radii of 30 kpc or more.

Stars in a stream trace a narrow bunch of orbits in the vicinity of
the orbit of the center of the mass, and are correlated in orbital
phase: they can all be traced back to a small volume (e.g., near
pericenters of the satellite orbit) where they were once bound to the
satellite.  Hence we expect a tight constraint on parameters of the
Galactic potential and the initial condition of the center of the mass
of the satellite (about a dozen parameters in total) by fitting the
individual proper motions of one hundred or more stars along a stream
since the fitting problem is over-constrained.

We propose to select bright horizontal and giant branch stars as
tracers of tidal debris of a halo satellite (which we take to be
either a dwarf galaxy or a globular cluster).  They are bright with
$M_V \le 0.75$mag, easily observable $V \le 18$mag within 20 kpc
from the Galactic center with GAIA.

\subsection{Field contamination}

There are numerous HB and GB stars in a satellite.  
Assume between $f=0.5\%$ and $f=50\%$ of the stars in the original
satellite are freed by two-body relaxation processes (as for a dense
globular cluster) and/or by tidal force (as for a fluffy dwarf galaxy).
Then the number of HB stars in a stream is
\beq
N_{\rm stream}= {f L_V \over (L_V/N_{\rm HB})} = 10^2-10^4,
\eeq
where we adopt one HB star per $L_V/N_{\rm HB}=(540\pm 40)L_\odot$ 
(Preston et al. 1991) for a globular cluster or a dwarf galaxy with 
a total luminosity $L_V=10^{5-7}L_\odot$.

In comparison, the entire metal-poor halo has a total luminosity of $(3 \pm
1)\times 10^7L_\odot$ and only about $N_{\rm halo}=(6\pm 2)\times
10^4$ HB stars (Kinman 1994).  These halo stars are very spread out in
velocity with a dispersion $\sigma_{\rm halo} \sim 3000\muasyr$ in
proper motion.  So
the number of field stars which happen to share the same proper motion
with a stream is
\beq
N_{\rm field}=N_{\rm halo} \left( {\sigma_{\rm stream} \over \sigma_{\rm
halo}} \right)^2 \le 70,
\eeq
where the dispersion of a stream $\sigma_{\rm stream}$
is generously set at $100\muasyr$, appropriate for a 
nearby massive dwarf galaxy with velocity dispersion of $10\kms$ at $20$kpc.  
The chance of confusing a HB/GB star in the field with in a stream
becomes even smaller if we select stars in a small patch 
of the sky with similar radial velocities and photometric parallax.
In fact Sgr was discovered just on the basis of radial velocity and
photometric parallax despite a dense foreground of bulge stars
(Ibata, Gilmore, \& Irwin 1994).  We conclude as far as 
identifying stars in a cold stream with GAIA is concerned the
contamination from field halo stars is likely not a serious problem.

We thank Amina Helmi for discussions and Tim de Zeeuw for helpful
comments on an earlier draft.


\clearpage

\onecolumn

\begin{figure}[]
\epsfysize=140mm
\leftline{\epsfbox{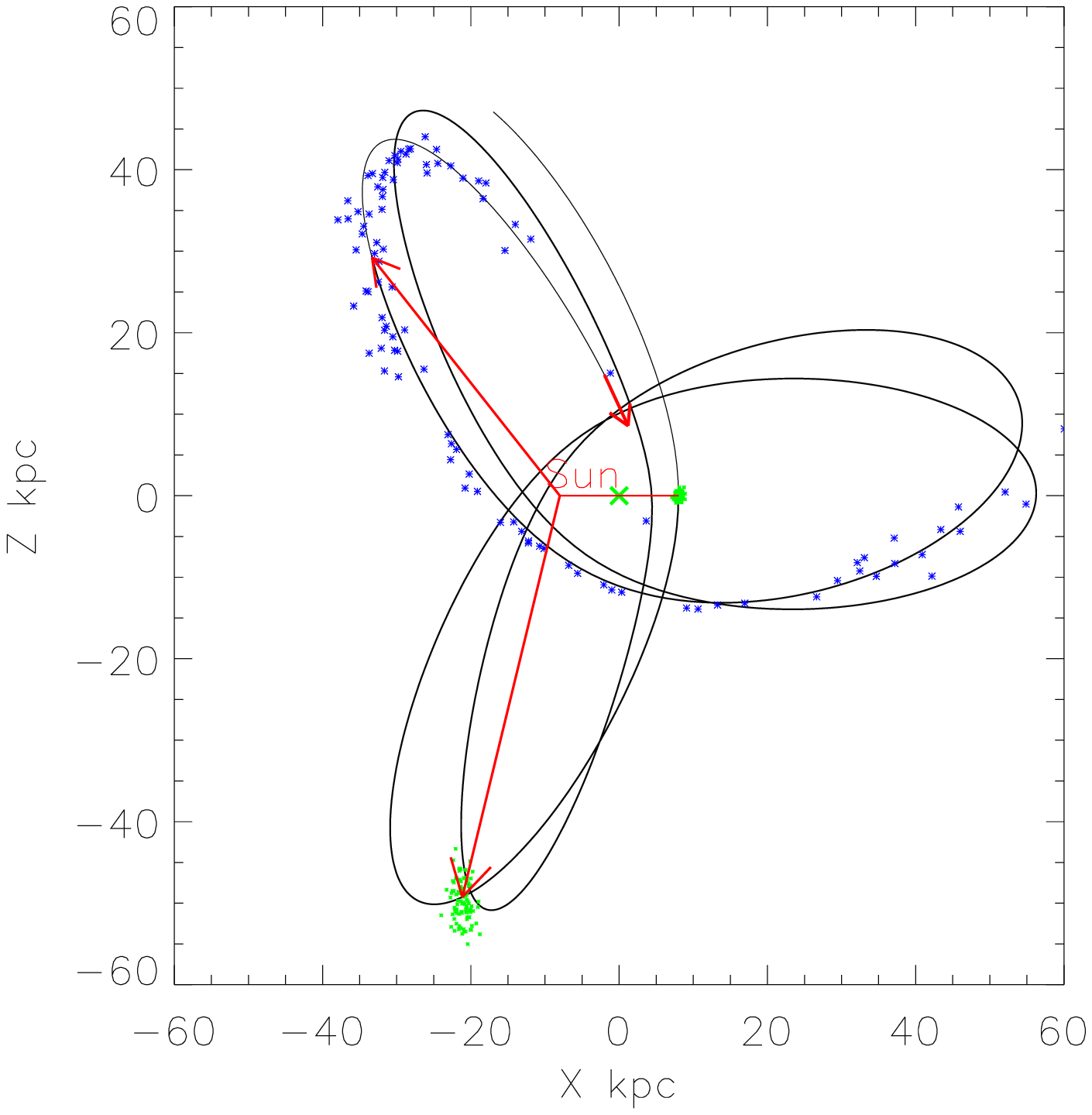}}
\caption{
Simulated orbit of the parent satellite and 
the released stream of about 100 giant stars in a potential model
with a growing disc $\Phi_G$.  
The Galactic center is marked by a cross.
We assume that the parent satellite is totally
disrupted 4 Gyrs ago at a pericenter $8$ kpc (Sun's mirror image point)
and moving south (down) with a velocity $380\kms$.  The stubby
debris tail shortly ($0.3$ Gyrs) after the disruption is also shown
together with the well-developed tidal tail 4 Gyrs later; 
the center of the satellite circles around like the arms of a clock,
with its position at different epoch is indicated by the long arrows.  
}
\end{figure}

\begin{figure}[]
\epsfysize=160mm
\center{\epsfbox{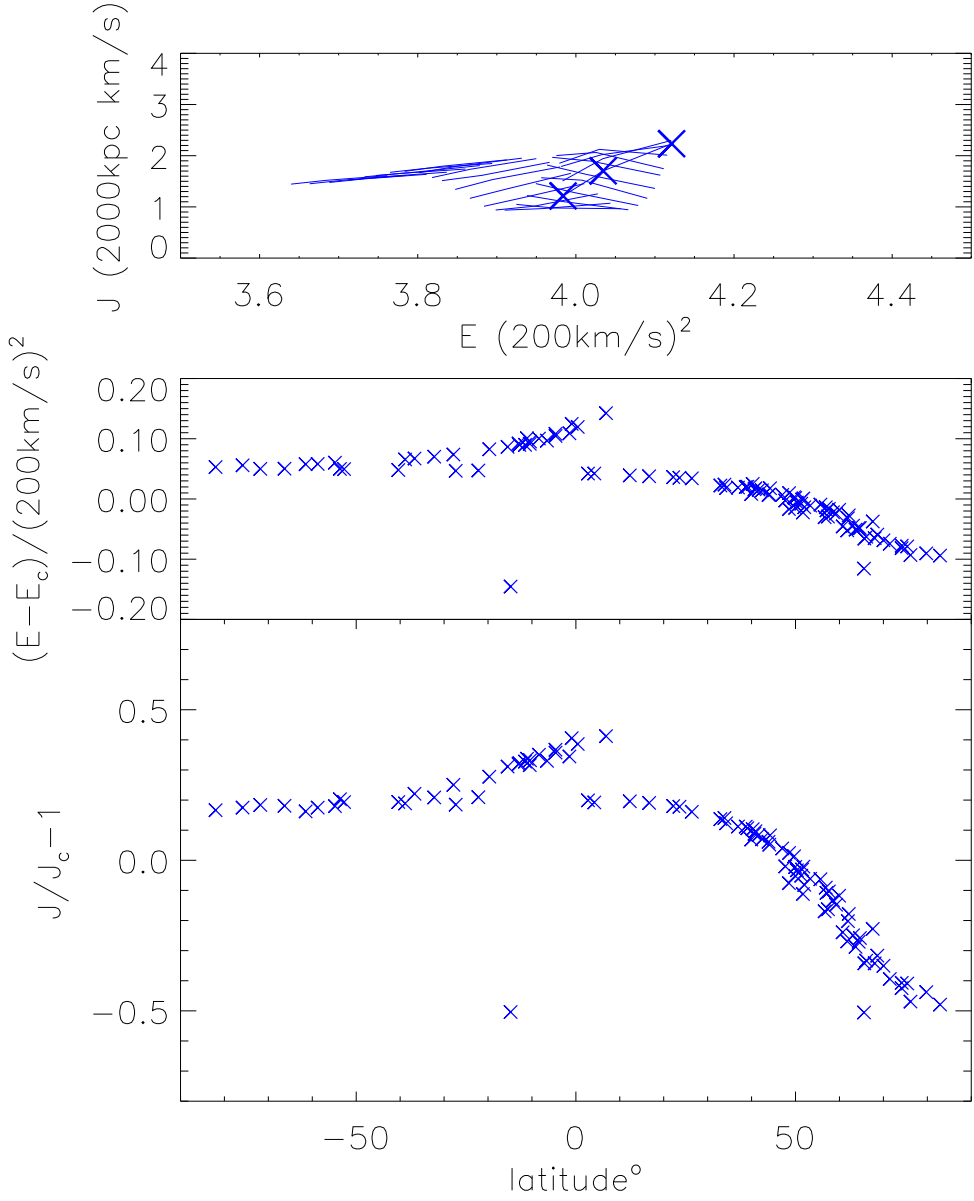}}
\caption{
The lower panels show the angular momentum and energy along the
stream simulated in the same time-varying potential as in Fig.1.  
$J_c$ and $E_c$ are the present values for the center of mass
of the stream.
The top panel shows how most ($\sim 70\%$) stars in a
stream evolve in the energy vs. angular momentum plane.  
Each short line segment is the distribution at an epoch; 
the present epoch is marked by the crosses.
We advance the stream in steps of about $0.15$ Gyr from 4 Gyrs ago to present.
}
\end{figure}

\begin{figure}[]
\epsfysize=200mm
\center{\epsfbox{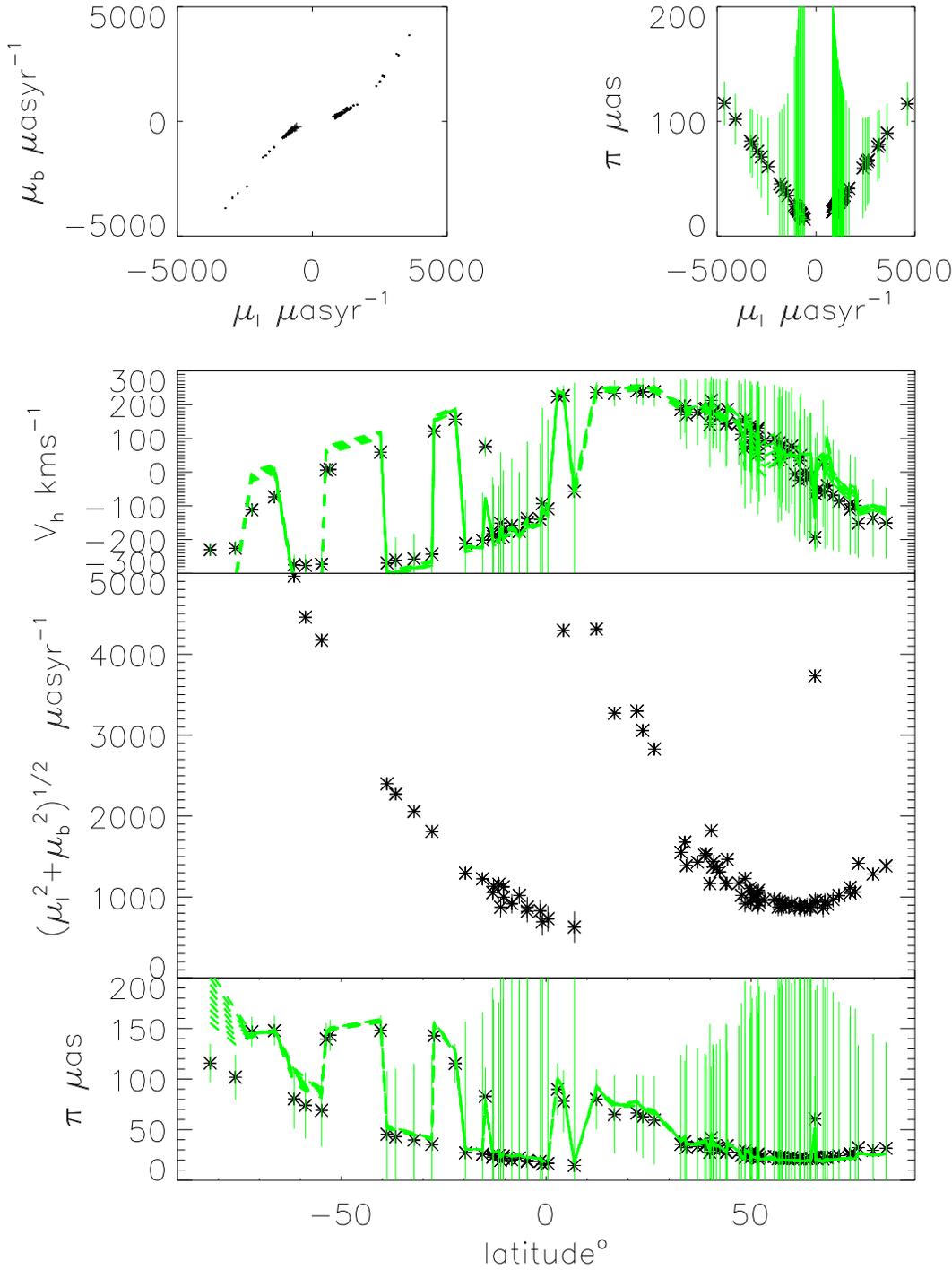}}
\caption{
Mock data of the stream convolved with GAIA accuracy for the potential model
in Fig. 1.
The two small panels at the top are the proper motion
vs. proper motion ($\mu_l$ vs. $\mu_b$) diagram and the (longitude)
proper motion vs. parallax ($\mu_l$ vs $\pi$) diagram; the linear
regression $|\mu_l| \sim 200 \kms/{\rm D\, kpc} \sim 40\pi$ 
manifests the reflex of the Galactic rotation of the Sun 
and the absence of any azimuthal rotation of a polar stream.  
The lower three panels show the heliocentric 
radial velocity, the proper motion, and the parallax across
the stream as functions of the latitude, together
with their error bars (thin vertical lines); the GAIA error bars for the
proper motion are typically much smaller than the size of the symbols.  
The narrow bands (heavily lines) 
show parallax and heliocentric velocity
predicted from GAIA proper motions, 
assuming a constant angular momentum $|{\bf J}|$ and
approximate energy $E$ across the stream; the width of the bands here are
computed from the initial spread of $E$ and ${\bf J}$.
}
\end{figure}

\begin{figure}[]
\epsfysize=160mm
\center{\epsfbox{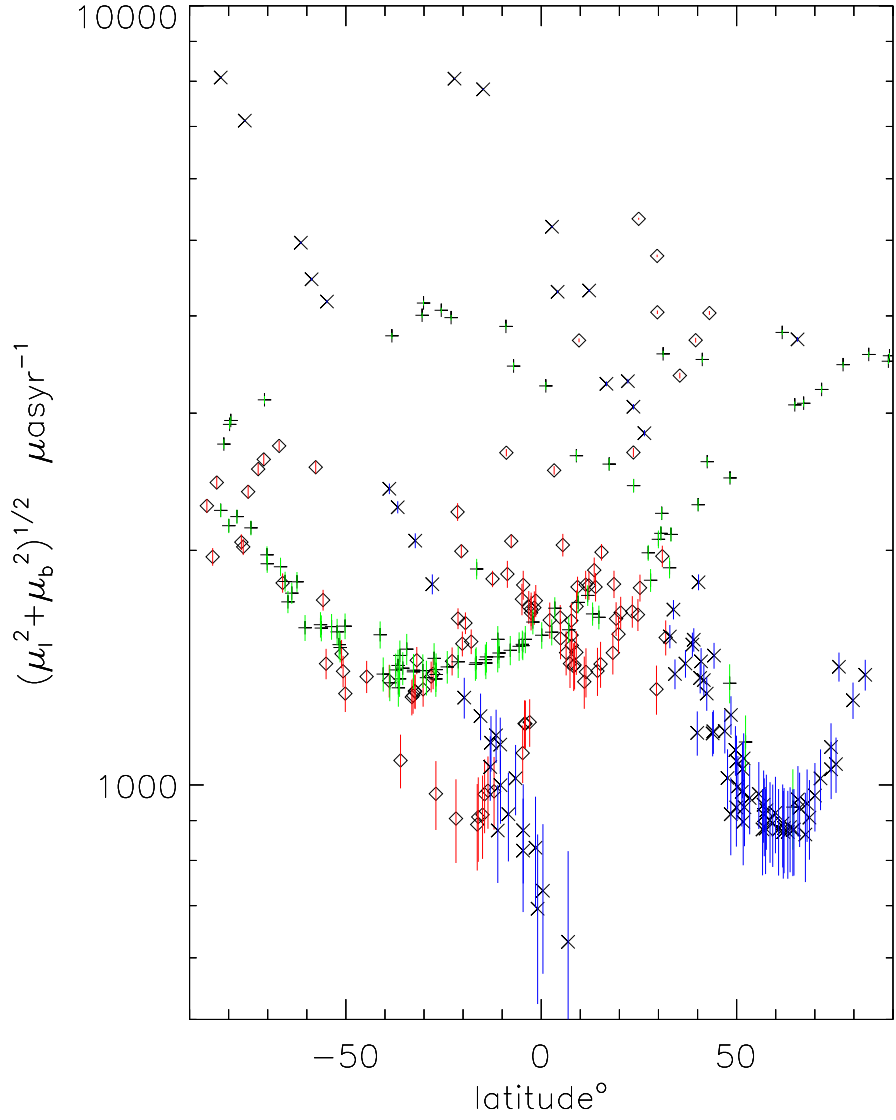}}
\caption{
Mock data for the norm of the proper motion vector across three streams
simulated with GAIA accuracy.  The three streams are generated
in three different evolution histories of the Galactic potential
(cf. Zhao et al. 1999).
Crosses are stars in a stream in 
the potential with a growing disc $\Phi_G$,
pluses are the disc-flapping potential $\Phi_F$.
diamonds are the potential with a massive perturber $\Phi_P$.
have the same meaning as in the previous figure.
}
\end{figure}

\end{document}